\documentclass[
  aip,
  amsmath,amssymb,
  preprint
]{revtex4-1}

\usepackage{graphicx}
\usepackage{amsmath,amssymb,amsthm}
\usepackage{physics}
\usepackage{booktabs}
\usepackage{array}
\usepackage{enumitem}
\usepackage{xcolor}
\usepackage{tikz}
\usetikzlibrary{patterns,calc,arrows.meta}
\usepackage{algpseudocode}

\usepackage{hyperref}

\newtheorem{theorem}{Theorem}[section]

\newtheorem{remark}{Remark}[section]

\begin{document}

\title{Quantum Mechanics in a Spherical Wedge: Complete Solution and Implications for Angular Momentum Theory}

\author{Mustafa Bakr}
\email{mustafa.bakr@physics.ox.ac.uk}
\affiliation{Clarendon Laboratory, Department of Physics, University of Oxford}

\author{Smain Amari}
% \affiliation{...} % add if you have one

\begin{abstract}
We solve the stationary Schr\"odinger equation for a particle confined to a three-dimensional spherical wedge---the region $\{(r,\theta,\phi): 0 \leq r \leq R,\, 0 \leq \theta \leq \pi,\, 0 \leq \phi \leq \Phi\}$ with Dirichlet boundary conditions on all surfaces. This exactly solvable constrained-domain model exhibits spectral reorganisation under symmetry-breaking boundary conditions and provides an operator-domain viewpoint on angular momentum quantisation. We obtain three main results. First, the stationary states are standing waves in the azimuthal coordinate and consequently are \emph{not} eigenstates of $\hat{L}_z$; we prove $\langle L_z \rangle = 0$ with $\Delta L_z = \hbar n_\phi\pi/\Phi \neq 0$, demonstrating that angular momentum projection becomes an observable with genuine quantum uncertainty rather than a good quantum number. Second, the effective azimuthal quantum number $\mu = n_\phi\pi/\Phi$ is generically non-integer, and square-integrability of the polar wavefunctions at both poles requires the angular eigenvalue parameter $\nu$ to satisfy $\nu - \mu \in \mathbb{Z}_{\geq 0}$. This regularity constraint yields a hierarchy: sectoral solutions ($\nu = \mu$, satisfying the first-order highest-weight condition) exist for any real $\mu > 0$, while tesseral and zonal solutions require integer steps, appearing only when $\mu$ itself is integer. Third, application to a Coulomb potential shows that the familiar integer angular momentum spectrum of hydrogen arises from the periodic identification $\phi \sim \phi + 2\pi$ that defines the full-sphere Hilbert space domain; modified boundary conditions yield a reorganised spectrum with non-integer effective angular momentum. The model clarifies the distinct roles of single-valuedness (selecting integer $m$ via azimuthal topology) and polar regularity (selecting integer $\ell \geq |m|$ via analytic constraints) in the standard quantisation of orbital angular momentum.
\end{abstract}

\maketitle

%##############################################################################
\section{Introduction}
\label{sec:intro}
%##############################################################################

The quantisation of orbital angular momentum is among the earliest and most celebrated results of quantum mechanics. The eigenvalues of $\hat{L}^2$ are $\hbar^2\ell(\ell+1)$ with $\ell \in \{0, 1, 2, \ldots\}$, and those of $\hat{L}_z$ are $\hbar m$ with $m \in \{-\ell, \ldots, \ell\}$. These integer quantum numbers emerge from the requirement that wavefunctions be single-valued under $2\pi$ rotations and regular at the poles of the sphere. The mathematical origin of these constraints is well understood: the configuration space of a particle on the unit sphere $S^2$ identifies the azimuthal coordinate $\phi$ with $\phi + 2\pi$, and the Hilbert space $L^2(S^2)$ consists of square-integrable functions on this compact manifold. The angular momentum operators are self-adjoint on this domain, and their spectra follow from representation theory of SO(3). This perspective is standard in textbooks~\cite{Messiah, Sakurai, CohenTannoudji}. A point worth emphasizing, though often left implicit, is that the familiar angular momentum spectrum emerges only after \emph{both} the differential expression and the operator domain are specified. The commutation relations $[\hat{L}_i, \hat{L}_j] = i\hbar\epsilon_{ijk}\hat{L}_k$ constrain the algebraic structure but admit both integer and half-integer representations; it is the domain---square-integrable functions on $S^2$ with periodic identification---that selects the integer orbital case.

Nevertheless, exactly solvable models that exhibit spectral reorganisation under modified boundary conditions can illuminate these constraints in ways that abstract arguments cannot. The present paper develops one such model: a quantum particle confined to a spherical wedge, the three-dimensional region bounded by a sphere and two half-planes meeting at the polar axis. This ``watermelon slice'' geometry---characterised by azimuthal extent $0 < \Phi < 2\pi$---replaces the periodic identification $\phi \sim \phi + 2\pi$ with Dirichlet boundary conditions at $\phi = 0$ and $\phi = \Phi$. Two-dimensional wedge domains are familiar from electrostatics and potential theory; the ``pie-shaped region'' appears in standard PDE texts as an exercise in separation of variables. The quantum billiard literature has extensively studied particles confined to two-dimensional domains of various shapes, including rectangles, ellipses, triangles, and chaotic geometries~\cite{Griffiths}. The three-dimensional spherical wedge, however, has received less systematic attention as a quantum-mechanical model, despite its exact solvability and the clarity with which it exhibits the interplay between boundary conditions and angular momentum quantisation.

The problem separates in spherical coordinates, yielding three ordinary differential equations. The azimuthal equation, subject to Dirichlet conditions, produces standing-wave solutions $\sin(n_\phi\pi\phi/\Phi)$ with effective quantum number $\mu = n_\phi\pi/\Phi$. The polar equation is the associated Legendre equation with non-integer order $\mu$; square-integrability at both poles constrains the separation constant. The radial equation involves spherical Bessel functions.

The analysis yields three principal results that we believe merit attention. \\ \textbf{1. Loss of $\hat{L}_z$ as a good quantum number.} The standing-wave structure of the azimuthal wavefunctions implies that stationary states are superpositions of counter-propagating angular momentum eigenstates $e^{\pm i\mu\phi}$. We prove that $\langle L_z \rangle = 0$ for all states while $\Delta L_z = \hbar\mu \neq 0$. The operator $\hat{L}_z = -i\hbar\partial/\partial\phi$ is not self-adjoint on the wedge Hilbert space (it maps functions satisfying Dirichlet conditions to functions violating them), and consequently the $z$-component of angular momentum exhibits genuine quantum uncertainty rather than definite eigenvalues. \\ \textbf{2. Sectoral-tesseral hierarchy under non-integer $\mu$.} The polar equation admits square-integrable solutions only when $\nu - \mu \in \mathbb{Z}_{\geq 0}$, where $\nu(\nu+1)$ is the angular eigenvalue. The ground state in each azimuthal sector is the sectoral solution $\nu = \mu$, for which the polar wavefunction $\Theta(\theta) \propto (\sin\theta)^\mu$ satisfies the first-order highest-weight condition $\hat{L}_+Y = 0$. This first-order character permits solutions for any real $\mu > 0$. Tesseral solutions ($\nu > \mu$) require the integer ladder $\nu = \mu + k$ and appear as excited polar states. When $\Phi = 2\pi/n$ for integer $n$, the ground-state $\mu = n/2$ can be half-integer---a value inaccessible in the full-sphere geometry. \\ \textbf{3. Spectral reorganisation for Coulomb potential.} Application to a hydrogen-like atom confined to a wedge (with no radial wall, only azimuthal confinement) yields energy levels $E = -13.6\,\text{eV}/(n_r + \nu + 1)^2$ with $\nu = n_\phi\pi/\Phi + k$. The standard hydrogen spectrum is recovered exactly when $\Phi = 2\pi$ with periodic boundary conditions; the wedge geometry accesses different points on the underlying energy-versus-angular-momentum relationship.

We emphasise that this paper presents an exactly solvable constrained-domain model, not a claim of ``new quantum mechanics.'' The mathematics---separation of variables, associated Legendre functions, spherical Bessel functions---is entirely standard. The contribution is pedagogical and conceptual: the model provides explicit, algebraically complete illustrations of how the specification of the Hilbert space domain affects the angular momentum spectrum. The paper is organised as follows. Section~\ref{sec:separation} sets up the problem and performs the separation of variables. Sections~\ref{sec:azimuthal}--\ref{sec:radial} solve the three resulting equations. Section~\ref{sec:angular_momentum} analyses the angular momentum properties. Section~\ref{sec:spectrum} presents the complete spectrum. Section~\ref{sec:hydrogen} applies the results to the hydrogen atom. Section~\ref{sec:conclusion} summarises the physical insights.

%##############################################################################
\section{Separation of Variables}
\label{sec:separation}
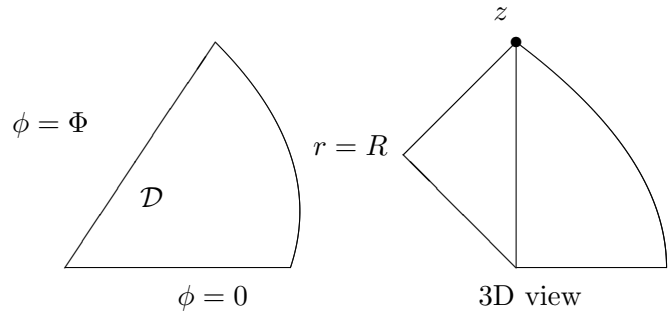
\begin{figure}[t]
\centering
\setlength{\unitlength}{1cm}
\begin{picture}(10,5)
% Draw wedge cross-section
\put(1,1){\line(1,0){3}}
\put(1,1){\line(2,3){2}}
\qbezier(4,1)(4.5,2.5)(3,4)
% Labels
\put(2.5,0.5){$\phi = 0$}
\put(0.3,2.8){$\phi = \Phi$}
\put(4.3,2.5){$r = R$}
\put(2,1.8){$\mathcal{D}$}
% 3D sketch
\put(7,1){\line(0,1){3}}
\put(7,1){\line(1,0){2}}
\put(7,1){\line(-1,1){1.5}}
\qbezier(9,1)(9,2.5)(7,4)
\qbezier(5.5,2.5)(6.2,3.2)(7,4)
\put(7,4){\circle*{0.15}}
\put(6.7,4.3){$z$}
\put(6.5,0.5){3D view}
\end{picture}
\caption{The spherical wedge geometry. Left: cross-section in a plane of constant $\theta$, showing the azimuthal extent $\Phi$. Right: three-dimensional ``watermelon slice'' region $\mathcal{D}$.}
\label{fig:geometry}
\end{figure}

%##############################################################################
%\subsection{The Domain and Boundary Conditions}
The spherical wedge is the three-dimensional region (Figure~\ref{fig:geometry})
\begin{equation}
\mathcal{D} = \{(r,\theta,\phi) : 0 \leq r \leq R, \; 0 \leq \theta \leq \pi, \; 0 \leq \phi \leq \Phi\}
\label{eq:domain}
\end{equation}
where $0 < \Phi < 2\pi$ is the azimuthal extent and $R > 0$ is the radial extent. The domain is bounded by a sphere of radius $R$ and two half-planes meeting at the polar axis ($z$-axis).

Consider the Schr\"odinger Equation~\cite{Schrodinger1926} a particle of mass $M$ confined to the region $\mathcal{D}$ defined in~\eqref{eq:domain}. The potential is
\begin{equation}
V(\mathbf{r}) = \begin{cases}
0 & \text{if } (r,\theta,\phi) \in \mathcal{D}, \\
+\infty & \text{otherwise}.
\end{cases}
\label{eq:potential}
\end{equation}
Inside $\mathcal{D}$, the time-independent Schr\"odinger equation is
\begin{equation}
-\frac{\hbar^2}{2M}\nabla^2\psi = E\psi.
\label{eq:schrodinger}
\end{equation}
The boundary conditions imposed by the infinite potential are:
\begin{align}
\psi(R, \theta, \phi) &= 0 \quad \text{for all } \theta, \phi, \label{eq:bc_radial}\\
\psi(r, \theta, 0) &= 0 \quad \text{for all } r, \theta, \label{eq:bc_phi0}\\
\psi(r, \theta, \Phi) &= 0 \quad \text{for all } r, \theta. \label{eq:bc_phiPhi}
\end{align}
In spherical coordinates $(r, \theta, \phi)$, the Laplacian is
\begin{equation}
\nabla^2 = \frac{1}{r^2}\frac{\partial}{\partial r}\left(r^2\frac{\partial}{\partial r}\right) + \frac{1}{r^2\sin\theta}\frac{\partial}{\partial\theta}\left(\sin\theta\frac{\partial}{\partial\theta}\right) + \frac{1}{r^2\sin^2\theta}\frac{\partial^2}{\partial\phi^2}.
\label{eq:laplacian}
\end{equation}
Substituting into~\eqref{eq:schrodinger} and multiplying by $-2Mr^2/\hbar^2$:
\begin{equation}
\frac{\partial}{\partial r}\left(r^2\frac{\partial\psi}{\partial r}\right) + \frac{1}{\sin\theta}\frac{\partial}{\partial\theta}\left(\sin\theta\frac{\partial\psi}{\partial\theta}\right) + \frac{1}{\sin^2\theta}\frac{\partial^2\psi}{\partial\phi^2} + \frac{2MEr^2}{\hbar^2}\psi = 0.
\label{eq:schrodinger_spherical}
\end{equation}
We seek solutions of the separable form
\begin{equation}
\psi(r,\theta,\phi) = \mathcal{R}(r)\Theta(\theta)\Phi(\phi).
\label{eq:ansatz}
\end{equation}
Substituting into~\eqref{eq:schrodinger_spherical} and dividing by $\mathcal{R}\Theta\Phi$:
\begin{equation}
\frac{1}{\mathcal{R}}\frac{d}{dr}\left(r^2\frac{d\mathcal{R}}{dr}\right) + \frac{2MEr^2}{\hbar^2} + \frac{1}{\Theta\sin\theta}\frac{d}{d\theta}\left(\sin\theta\frac{d\Theta}{d\theta}\right) + \frac{1}{\Phi\sin^2\theta}\frac{d^2\Phi}{d\phi^2} = 0.
\label{eq:separated_1}
\end{equation}
Multiply through by $\sin^2\theta$ and rearrange:
\begin{equation}
\sin^2\theta\left[\frac{1}{\mathcal{R}}\frac{d}{dr}\left(r^2\frac{d\mathcal{R}}{dr}\right) + \frac{2MEr^2}{\hbar^2}\right] + \frac{\sin\theta}{\Theta}\frac{d}{d\theta}\left(\sin\theta\frac{d\Theta}{d\theta}\right) = -\frac{1}{\Phi}\frac{d^2\Phi}{d\phi^2}.
\label{eq:separated_2}
\end{equation}
The left side depends on $r$ and $\theta$; the right side depends only on $\phi$. For this equation to hold for all $(r,\theta,\phi)$, both sides must equal a constant, which we call $\mu^2$:
\begin{equation}
\frac{d^2\Phi}{d\phi^2} = -\mu^2\Phi.
\label{eq:azimuthal_ode}
\end{equation}
The remaining equation, after dividing by $\sin^2\theta$, is:
\begin{equation}
\frac{1}{\mathcal{R}}\frac{d}{dr}\left(r^2\frac{d\mathcal{R}}{dr}\right) + \frac{2MEr^2}{\hbar^2} + \frac{1}{\Theta\sin\theta}\frac{d}{d\theta}\left(\sin\theta\frac{d\Theta}{d\theta}\right) - \frac{\mu^2}{\sin^2\theta} = 0.
\label{eq:separated_3}
\end{equation}
The first two terms depend only on $r$; the last two depend only on $\theta$. Setting both equal to a constant $\lambda$:
\begin{align}
\frac{1}{\mathcal{R}}\frac{d}{dr}\left(r^2\frac{d\mathcal{R}}{dr}\right) + \frac{2MEr^2}{\hbar^2} &= \lambda, \label{eq:radial_separation}\\
\frac{1}{\Theta\sin\theta}\frac{d}{d\theta}\left(\sin\theta\frac{d\Theta}{d\theta}\right) - \frac{\mu^2}{\sin^2\theta} &= -\lambda. \label{eq:polar_separation}
\end{align}
We have obtained three ordinary differential equations: the azimuthal equation~\eqref{eq:azimuthal_ode}, the polar equation~\eqref{eq:polar_separation}, and the radial equation~\eqref{eq:radial_separation}. We now solve each in turn.

%#############################################################################
\section{The Azimuthal Equation}
\label{sec:azimuthal}
%#############################################################################
The azimuthal equation~\eqref{eq:azimuthal_ode} is
\begin{equation}
\frac{d^2\Phi}{d\phi^2} + \mu^2\Phi = 0.
\label{eq:azimuthal_ode_2}
\end{equation}
This is the simple harmonic oscillator equation with general solution
\begin{equation}
\Phi(\phi) = A\cos(\mu\phi) + B\sin(\mu\phi),
\label{eq:azimuthal_general}
\end{equation}
where $A$ and $B$ are constants to be determined by boundary conditions. The boundary conditions~\eqref{eq:bc_phi0} and~\eqref{eq:bc_phiPhi} require:
\begin{align}
\Phi(0) &= A = 0, \label{eq:bc_phi0_applied}\\
\Phi(\Phi) &= B\sin(\mu\Phi) = 0. \label{eq:bc_phiPhi_applied}
\end{align}
From~\eqref{eq:bc_phi0_applied}, the cosine term vanishes: $A = 0$. From~\eqref{eq:bc_phiPhi_applied}, either $B = 0$ (trivial solution) or $\sin(\mu\Phi) = 0$. The latter requires
\begin{equation}
\mu\Phi = n_\phi\pi, \quad n_\phi = 1, 2, 3, \ldots
\label{eq:mu_quantisation}
\end{equation}
We exclude $n_\phi = 0$ because it gives $\Phi(\phi) = 0$ identically, and we exclude negative $n_\phi$ because $\sin(-x) = -\sin(x)$ merely changes the sign of $B$, which is absorbed into normalisation.

Consider the quantized values and normalization, the allowed values of the separation constant are
\begin{equation}
\mu_{n_\phi} = \frac{n_\phi\pi}{\Phi}, \quad n_\phi = 1, 2, 3, \ldots
\label{eq:mu_values}
\end{equation}
The azimuthal wavefunctions are
\begin{equation}
\Phi_{n_\phi}(\phi) = B\sin\left(\frac{n_\phi\pi\phi}{\Phi}\right).
\label{eq:azimuthal_unnormalised}
\end{equation}
Normalisation requires $\int_0^\Phi |\Phi_{n_\phi}|^2 d\phi = 1$:
\begin{equation}
|B|^2\int_0^\Phi \sin^2\left(\frac{n_\phi\pi\phi}{\Phi}\right)d\phi = |B|^2 \cdot \frac{\Phi}{2} = 1.
\label{eq:azimuthal_normalisation}
\end{equation}
Therefore $|B| = \sqrt{2/\Phi}$, and the normalised azimuthal wavefunctions are
\begin{equation}
\boxed{\Phi_{n_\phi}(\phi) = \sqrt{\frac{2}{\Phi}}\sin\left(\frac{n_\phi\pi\phi}{\Phi}\right), \quad n_\phi = 1, 2, 3, \ldots}
\label{eq:azimuthal_normalised}
\end{equation}

\subsection{Non-Integer Effective Quantum Number}
For a general wedge angle $\Phi$, the effective azimuthal quantum number $\mu = n_\phi\pi/\Phi$ is non-integer. Some examples:
\begin{center}
\begin{tabular}{lcccc}
\hline
Geometry & $\Phi$ & $\mu_1$ & $\mu_2$ & $\mu_3$ \\
\hline
Hemisphere & $\pi$ & 1 & 2 & 3 \\
Third-sphere & $2\pi/3$ & $3/2$ & 3 & $9/2$ \\
Quarter-sphere & $\pi/2$ & 2 & 4 & 6 \\
Golden wedge & $2\pi/\varphi$ ($\varphi = $ golden ratio) & $\varphi/2$ & $\varphi$ & $3\varphi/2$ \\
\hline
\end{tabular}
\end{center}
The third-sphere case ($\Phi = 2\pi/3$) is particularly notable: the ground state has $\mu_1 = 3/2$, a half-integer that never appears in the standard theory of orbital angular momentum.

%#############################################################################
\section{The Polar Equation}
\label{sec:polar}
%#############################################################################
The polar equation~\eqref{eq:polar_separation} can be written as
\begin{equation}
\frac{1}{\sin\theta}\frac{d}{d\theta}\left(\sin\theta\frac{d\Theta}{d\theta}\right) + \left[\lambda - \frac{\mu^2}{\sin^2\theta}\right]\Theta = 0.
\label{eq:polar_ode}
\end{equation}
This is the associated Legendre equation. With the substitution $x = \cos\theta$ (so $dx = -\sin\theta\,d\theta$ and $\sin^2\theta = 1 - x^2$), it becomes
\begin{equation}
\frac{d}{dx}\left[(1-x^2)\frac{d\Theta}{dx}\right] + \left[\lambda - \frac{\mu^2}{1-x^2}\right]\Theta = 0.
\label{eq:legendre_ode}
\end{equation}

\subsection{Singular Points and Regularity}
The points $x = \pm 1$ (corresponding to $\theta = 0$ and $\theta = \pi$, the poles) are regular singular points of equation~\eqref{eq:legendre_ode}. Near $x = 1$, Frobenius analysis shows that the two linearly independent solutions behave as
\begin{equation}
\Theta \sim (1-x)^{\pm\mu/2} \quad \text{as } x \to 1^-.
\label{eq:behaviour_x1}
\end{equation}
We require solutions that are \emph{square-integrable} under the natural measure:
\begin{equation}
\int_0^\pi |\Theta(\theta)|^2 \sin\theta\, d\theta < \infty.
\label{eq:square_integrability}
\end{equation}
Near $\theta = 0$, we have $\sin\theta \approx \theta$ and $(1-\cos\theta) \approx \theta^2/2$, so $(1-x)^{\mu/2} \sim \theta^\mu$. The integrand behaves as $\theta^{2\mu} \cdot \theta = \theta^{2\mu+1}$, which is integrable for $\mu > -1$. In contrast, $(1-x)^{-\mu/2} \sim \theta^{-\mu}$ gives $\theta^{-2\mu+1}$, which diverges for $\mu > 1$. For $\mu > 0$, we must therefore select the solution $(1-x)^{+\mu/2}$ near $x = 1$. Similarly, near $x = -1$ (the south pole):
\begin{equation}
\Theta \sim (1+x)^{\pm\mu/2} \quad \text{as } x \to -1^+.
\label{eq:behaviour_xm1}
\end{equation}
Again, square-integrability requires the solution $(1+x)^{+\mu/2}$ near $x = -1$.

The requirement that $\Theta$ be square-integrable at \emph{both} poles places a constraint on the separation constant $\lambda$. The two linearly independent solutions of the associated Legendre equation are the Ferrers functions $P_\nu^\mu(x)$ and $Q_\nu^\mu(x)$~\cite{DLMF}. For real $\mu > 0$ and $x \in (-1, 1)$: The function $P_\nu^\mu(x)$ (Ferrers function of the first kind) behaves as $(1-x^2)^{\mu/2}$ times a function regular at both endpoints when $\nu - \mu$ is a non-negative integer. Otherwise, it has a logarithmic singularity at $x = -1$. The function $Q_\nu^\mu(x)$ (Ferrers function of the second kind) always diverges at $x = \pm 1$ and is therefore excluded on integrability grounds for any values of $\nu$ and $\mu$. The general solution of the associated Legendre equation is $A P_\nu^\mu + B Q_\nu^\mu$; square-integrability forces $B = 0$.

Specifically, the DLMF~\cite{DLMF} (§14.3 and §14.8) establishes that for $\mu > 0$, the Ferrers function $P_\nu^\mu(x)$ is bounded on the closed interval $[-1, 1]$ if and only if
\begin{equation}
\nu - \mu = k \quad \text{for some } k \in \{0, 1, 2, 3, \ldots\}.
\label{eq:regularity_condition}
\end{equation}
When this condition holds, $P_\nu^\mu(\cos\theta)$ is square-integrable under~\eqref{eq:square_integrability}. When $\nu - \mu$ is not a non-negative integer, the would-be solution $P_\nu^\mu(x)$ diverges as $(1+x)^{-\mu}$ near $x = -1$, violating square-integrability. In other words, the allowed values are $\nu = \mu, \mu+1, \mu+2, \ldots$, with $\lambda = \nu(\nu+1)$.

\subsection{Proof of the Sectoral Solution}
The case $k = 0$ ($\nu = \mu$) admits an explicit solution that we now derive. We seek a solution of the form $\Theta(\theta) = (\sin\theta)^\mu$.

\textbf{Step 1: Compute the derivatives.}
\begin{align}
\frac{d\Theta}{d\theta} &= \mu(\sin\theta)^{\mu-1}\cos\theta, \\
\sin\theta\frac{d\Theta}{d\theta} &= \mu(\sin\theta)^\mu\cos\theta.
\end{align}

\textbf{Step 2: Compute the second term.}
\begin{align}
\frac{d}{d\theta}\left(\sin\theta\frac{d\Theta}{d\theta}\right) &= \frac{d}{d\theta}\left[\mu(\sin\theta)^\mu\cos\theta\right] \nonumber\\
&= \mu\left[\mu(\sin\theta)^{\mu-1}\cos^2\theta - (\sin\theta)^{\mu+1}\right] \nonumber\\
&= \mu(\sin\theta)^{\mu-1}\left[\mu\cos^2\theta - \sin^2\theta\right].
\end{align}

\textbf{Step 3: Substitute into the polar equation.}
Dividing by $\sin\theta$ and adding the other terms:
\begin{align}
&\frac{1}{\sin\theta}\frac{d}{d\theta}\left(\sin\theta\frac{d\Theta}{d\theta}\right) + \left[\lambda - \frac{\mu^2}{\sin^2\theta}\right]\Theta \nonumber\\
&= \mu(\sin\theta)^{\mu-2}\left[\mu\cos^2\theta - \sin^2\theta\right] + \lambda(\sin\theta)^\mu - \mu^2(\sin\theta)^{\mu-2} \nonumber\\
&= (\sin\theta)^{\mu-2}\left[\mu^2\cos^2\theta - \mu\sin^2\theta + \lambda\sin^2\theta - \mu^2\right] \nonumber\\
&= (\sin\theta)^{\mu-2}\left[\mu^2(\cos^2\theta - 1) + (\lambda - \mu)\sin^2\theta\right] \nonumber\\
&= (\sin\theta)^{\mu-2}\left[-\mu^2\sin^2\theta + (\lambda - \mu)\sin^2\theta\right] \nonumber\\
&= (\sin\theta)^\mu\left[\lambda - \mu - \mu^2\right].
\end{align}

\textbf{Step 4: Determine $\lambda$.}
For this to vanish, we need $\lambda = \mu(\mu + 1)$. This confirms that $\Theta(\theta) = (\sin\theta)^\mu$ satisfies the polar equation with $\lambda = \mu(\mu+1)$, i.e., $\nu = \mu$.

\textbf{Step 5: Check regularity.}
For $\mu > 0$, we have $(\sin\theta)^\mu \to 0$ as $\theta \to 0$ or $\theta \to \pi$. The solution is bounded (in fact, zero) at both poles.

For $k \geq 1$, the solutions are associated Legendre functions $P_\nu^\mu(\cos\theta)$ with $\nu = \mu + k$. These have $k$ nodes in the interval $(0, \pi)$ and are more complex to write explicitly. The key point is that they exist only when $k$ is a non-negative integer.

\subsection{Normalisation}
The polar wavefunctions must satisfy $\int_0^\pi |\Theta|^2 \sin\theta\, d\theta = 1$. For the sectoral case $\Theta(\theta) = N(\sin\theta)^\mu$:
\begin{align}
|N|^2\int_0^\pi (\sin\theta)^{2\mu}\sin\theta\,d\theta &= |N|^2\int_0^\pi (\sin\theta)^{2\mu+1}d\theta \nonumber\\
&= |N|^2 \cdot \frac{\sqrt{\pi}\,\Gamma(\mu+1)}{\Gamma(\mu + 3/2)} = 1.
\end{align}
Therefore
\begin{equation}
\Theta_{\mu,0}(\theta) = \sqrt{\frac{\Gamma(\mu+3/2)}{\sqrt{\pi}\,\Gamma(\mu+1)}}(\sin\theta)^\mu.
\label{eq:polar_normalised}
\end{equation}

%#############################################################################
\section{The Radial Equation}
\label{sec:radial}
%#############################################################################
With $\lambda = \nu(\nu+1)$, the radial equation~\eqref{eq:radial_separation} becomes
\begin{equation}
\frac{d}{dr}\left(r^2\frac{d\mathcal{R}}{dr}\right) + \left[\frac{2MEr^2}{\hbar^2} - \nu(\nu+1)\right]\mathcal{R} = 0.
\label{eq:radial_ode}
\end{equation}
Define $k^2 = 2ME/\hbar^2$ (assuming $E > 0$). Expanding the derivative:
\begin{equation}
r^2\frac{d^2\mathcal{R}}{dr^2} + 2r\frac{d\mathcal{R}}{dr} + \left[k^2r^2 - \nu(\nu+1)\right]\mathcal{R} = 0.
\label{eq:radial_ode_expanded}
\end{equation}

With the substitution $\rho = kr$, this becomes the spherical Bessel equation:
\begin{equation}
\rho^2\frac{d^2\mathcal{R}}{d\rho^2} + 2\rho\frac{d\mathcal{R}}{d\rho} + \left[\rho^2 - \nu(\nu+1)\right]\mathcal{R} = 0.
\label{eq:bessel_ode}
\end{equation}
The general solution is
\begin{equation}
\mathcal{R}(r) = Aj_\nu(kr) + By_\nu(kr),
\label{eq:radial_general}
\end{equation}
where $j_\nu$ and $y_\nu$ are spherical Bessel functions of the first and second kind.

\subsection{Boundary Conditions}
\textbf{At $r = 0$:} The spherical Bessel function of the second kind diverges: $y_\nu(\rho) \to -\infty$ as $\rho \to 0$ for all $\nu \geq 0$. Therefore $B = 0$.
\textbf{At $r = R$:} The boundary condition~\eqref{eq:bc_radial} requires $\mathcal{R}(R) = 0$, i.e.,
\begin{equation}
j_\nu(kR) = 0.
\label{eq:radial_bc}
\end{equation}
Let $\chi_{n_r,\nu}$ denote the $n_r$-th positive zero of $j_\nu(x)$, where $n_r = 1, 2, 3, \ldots$. Then
\begin{equation}
kR = \chi_{n_r,\nu} \quad \Rightarrow \quad k = \frac{\chi_{n_r,\nu}}{R}.
\label{eq:k_quantisation}
\end{equation}

\subsection{Energy Eigenvalues}
From $E = \hbar^2k^2/(2M)$:
\begin{equation}
\boxed{E_{n_r,\nu} = \frac{\hbar^2\chi_{n_r,\nu}^2}{2MR^2}}
\label{eq:energy_eigenvalue}
\end{equation}
where $\nu = \mu + k = n_\phi\pi/\Phi + k$ for $k = 0, 1, 2, \ldots$ and $n_\phi = 1, 2, 3, \ldots$.

\subsection{Behaviour Near the Origin for Small $\nu$}
For wedges wider than a hemisphere ($\Phi > \pi$), the ground-state sectoral solution has $\nu = \mu = \pi/\Phi < 1$. As $\Phi \to 2\pi$, this effective angular momentum approaches $\nu \to 1/2$. It is natural to ask whether the wavefunction becomes singular at $r = 0$ for small $\nu$. The spherical Bessel function of the first kind has the small-argument behaviour~\cite{DLMF}:
\begin{equation}
j_\nu(\rho) \sim \frac{\rho^\nu}{(2\nu+1)!!} \quad \text{as } \rho \to 0,
\label{eq:bessel_small_arg}
\end{equation}
so the radial wavefunction behaves as $\mathcal{R}(r) \sim r^\nu$ near the origin. For any $\nu > 0$, no matter how small, the wavefunction \emph{vanishes} at $r = 0$. There is no singularity in the wavefunction itself. However, for $0 < \nu < 1$, the wavefunction vanishes more slowly than for standard $\ell \geq 1$ states, leading to enhanced probability density near the origin compared to $p$, $d$, or $f$ orbitals.

The probability density $|\psi|^2 \propto r^{2\nu}$ remains integrable: the radial integral $\int_0^R r^{2\nu} \cdot r^2\,dr = \int_0^R r^{2\nu+2}\,dr$ converges for $\nu > -3/2$. The expectation value of kinetic energy, which involves $\int |\nabla\psi|^2 r^2\,dr \sim \int r^{2(\nu-1)} \cdot r^2\,dr = \int r^{2\nu}\,dr$, also converges for $\nu > -1/2$. Therefore, states with small $\nu$ are well-behaved quantum mechanically---bounded, normalisable, and with finite energy. The physical consequence is simply that particles in wide-angle wedges have higher probability density near the origin than their counterparts in narrow wedges or the full sphere.

\begin{remark}
This behaviour differs qualitatively from the electromagnetic case~\cite{bakrsphere, bakrcylinder1, bakrcylinder2}. For TM modes in a conducting wedge cavity, the electric field components can scale as $r^{\nu-1}$, which diverges as $r \to 0$ when $\nu < 1$. In the electromagnetic problem, this singularity is resolved by the finite conductivity of real metals or by the idealisation of a ``thin fin'' geometry. In quantum mechanics, however, the relevant quantity is the wavefunction $\psi \sim r^\nu$, not its gradient, and no such singularity arises.
\end{remark}
%#############################################################################
\section{Angular Momentum Properties}
\label{sec:angular_momentum}
%#############################################################################
The $z$-component of angular momentum is represented by the operator
\begin{equation}
\hat{L}_z = -i\hbar\frac{\partial}{\partial\phi}.
\label{eq:Lz_operator}
\end{equation}
In the standard treatment of angular momentum (full sphere with periodic boundary conditions), the eigenfunctions of $\hat{L}_z$ are $e^{im\phi}$ with integer $m$, and the eigenvalue is $\hbar m$.

\subsection{Wavefunctions Are Not $\hat{L}_z$ Eigenstates}
We now prove that the azimuthal wavefunctions~\eqref{eq:azimuthal_normalised} are \emph{not} eigenfunctions of $\hat{L}_z$. Acting with $\hat{L}_z$ on $\Phi_{n_\phi}$:
\begin{align}
\hat{L}_z\Phi_{n_\phi}(\phi) &= -i\hbar\frac{\partial}{\partial\phi}\left[\sqrt{\frac{2}{\Phi}}\sin\left(\frac{n_\phi\pi\phi}{\Phi}\right)\right] \nonumber\\
&= -i\hbar \cdot \frac{n_\phi\pi}{\Phi} \cdot \sqrt{\frac{2}{\Phi}}\cos\left(\frac{n_\phi\pi\phi}{\Phi}\right) \nonumber\\
&= -i\hbar\mu\sqrt{\frac{2}{\Phi}}\cos(\mu\phi).
\label{eq:Lz_action}
\end{align}
If $\Phi_{n_\phi}$ were an eigenfunction, the result would be proportional to $\Phi_{n_\phi}$ itself, i.e., to $\sin(\mu\phi)$. But $\cos(\mu\phi) \neq c\sin(\mu\phi)$ for any constant $c$. Therefore:
\begin{theorem}
The azimuthal wavefunctions of the spherical wedge are not eigenstates of $\hat{L}_z$.
\end{theorem}

The operator $\hat{L}_z$ generates rotations about the $z$-axis:
\begin{equation}
e^{-i\hat{L}_z\delta\phi/\hbar}\psi(\phi) = \psi(\phi + \delta\phi).
\label{eq:rotation_generator}
\end{equation}
For a state satisfying Dirichlet conditions at $\phi = 0$ and $\phi = \Phi$, an infinitesimal rotation produces a state that \emph{violates} the boundary conditions. The rotation operator does not preserve the domain of allowed states. Consequently, $\hat{L}_z$ does not commute with the Hamiltonian-plus-boundary-conditions system, and $L_z$ is not a conserved quantity.

\subsection{Expectation Value of $L_z$}
Although $\Phi_{n_\phi}$ is not an eigenstate of $\hat{L}_z$, we can compute the expectation value:
\begin{align}
\langle L_z \rangle &= \int_0^\Phi \Phi_{n_\phi}^*(\phi)\,\hat{L}_z\,\Phi_{n_\phi}(\phi)\,d\phi \nonumber\\
&= \frac{2}{\Phi}\int_0^\Phi \sin(\mu\phi) \cdot (-i\hbar\mu)\cos(\mu\phi)\,d\phi \nonumber\\
&= \frac{-i\hbar\mu}{\Phi}\int_0^\Phi \sin(2\mu\phi)\,d\phi \nonumber\\
&= \frac{-i\hbar\mu}{\Phi}\left[-\frac{\cos(2\mu\phi)}{2\mu}\right]_0^\Phi \nonumber\\
&= \frac{-i\hbar\mu}{\Phi} \cdot \frac{1 - \cos(2n_\phi\pi)}{2\mu} \nonumber\\
&= \frac{-i\hbar}{2\Phi}(1 - 1) = 0.
\label{eq:Lz_expectation}
\end{align}
\begin{theorem}
For all azimuthal states of the spherical wedge, $\langle L_z \rangle = 0$.
\end{theorem}

The expectation Value of $L_z^2$ is:
\begin{align}
\langle L_z^2 \rangle &= \int_0^\Phi \Phi_{n_\phi}^*(\phi)\,\hat{L}_z^2\,\Phi_{n_\phi}(\phi)\,d\phi \nonumber\\
&= \frac{2}{\Phi}\int_0^\Phi \sin(\mu\phi) \cdot (-\hbar^2)\frac{\partial^2}{\partial\phi^2}\sin(\mu\phi)\,d\phi \nonumber\\
&= \frac{2}{\Phi}\int_0^\Phi \sin(\mu\phi) \cdot (\hbar^2\mu^2)\sin(\mu\phi)\,d\phi \nonumber\\
&= \frac{2\hbar^2\mu^2}{\Phi}\int_0^\Phi \sin^2(\mu\phi)\,d\phi \nonumber\\
&= \frac{2\hbar^2\mu^2}{\Phi} \cdot \frac{\Phi}{2} = \hbar^2\mu^2.
\label{eq:Lz2_expectation}
\end{align}

Looking at the Uncertainty in $L_z$, the variance is
\begin{equation}
(\Delta L_z)^2 = \langle L_z^2 \rangle - \langle L_z \rangle^2 = \hbar^2\mu^2 - 0 = \hbar^2\mu^2.
\label{eq:Lz_variance}
\end{equation}
Therefore the uncertainty is
\begin{equation}
\boxed{\Delta L_z = \hbar\mu = \frac{n_\phi\pi\hbar}{\Phi}}
\label{eq:Lz_uncertainty}
\end{equation}
\begin{theorem}
The uncertainty in $L_z$ for a spherical wedge state is $\Delta L_z = \hbar n_\phi\pi/\Phi$, which is non-zero for all states.
\end{theorem}
The standing-wave structure explains these results. The sine function can be written as
\begin{equation}
\sin(\mu\phi) = \frac{e^{i\mu\phi} - e^{-i\mu\phi}}{2i}.
\label{eq:standing_wave_decomposition}
\end{equation}
The travelling waves $e^{\pm i\mu\phi}$ would be eigenstates of $\hat{L}_z$ with eigenvalues $\pm\hbar\mu$, but they individually violate the boundary conditions (neither vanishes at $\phi = 0$). The physical state is an equal superposition of both, with amplitudes $\pm 1/(2i)$. The probabilities are $|1/(2i)|^2 = 1/4$ each, but since we measure $+\hbar\mu$ or $-\hbar\mu$ with equal probability, the mean is zero:
\begin{equation}
\langle L_z \rangle = \frac{1}{2}(+\hbar\mu) + \frac{1}{2}(-\hbar\mu) = 0.
\end{equation}
The mean square is
\begin{equation}
\langle L_z^2 \rangle = \frac{1}{2}(\hbar\mu)^2 + \frac{1}{2}(-\hbar\mu)^2 = \hbar^2\mu^2,
\end{equation}
confirming our earlier calculation.

%#############################################################################
\section{Complete Energy Spectrum}
\label{sec:spectrum}
%#############################################################################
The complete set of quantum numbers for the spherical wedge is:
$n_\phi = 1, 2, 3, \ldots$ (azimuthal quantum number), with $\mu = n_\phi\pi/\Phi$.
$k = 0, 1, 2, \ldots$ (polar quantum number), with $\nu = \mu + k$.
$n_r = 1, 2, 3, \ldots$ (radial quantum number).
The energy eigenvalues are
\begin{equation}
\boxed{E_{n_r, k, n_\phi} = \frac{\hbar^2}{2MR^2}\chi_{n_r,\nu}^2, \quad \text{where } \nu = \frac{n_\phi\pi}{\Phi} + k}
\label{eq:full_spectrum}
\end{equation}

The ground state has $n_\phi = 1$, $k = 0$, $n_r = 1$:
\begin{equation}
E_{\text{ground}} = \frac{\hbar^2}{2MR^2}\chi_{1,\mu_1}^2, \quad \mu_1 = \frac{\pi}{\Phi}.
\label{eq:ground_state}
\end{equation}
For the hemisphere ($\Phi = \pi$), $\mu_1 = 1$ and $\chi_{1,1} \approx 4.493$, giving $E_{\text{ground}} \approx 20.2\hbar^2/(2MR^2)$.
For comparison, the full sphere has a ground state with $\ell = 0$, $\chi_{1,0} = \pi$, giving $E_{\text{ground}} = \pi^2\hbar^2/(2MR^2) \approx 9.87\hbar^2/(2MR^2)$.
The wedge confinement roughly doubles the ground-state energy. Table~\ref{tab:spectrum} shows the lowest energy levels for several wedge geometries.
\begin{table}[t]
\centering
\caption{Lowest energy levels $E \times 2MR^2/\hbar^2$ for various wedge angles.}
\label{tab:spectrum}
\begin{tabular}{lccccc}
\hline
& Full sphere & Hemisphere & Third-sphere & Quarter-sphere \\
& $\Phi = 2\pi$ & $\Phi = \pi$ & $\Phi = 2\pi/3$ & $\Phi = \pi/2$ \\
\hline
Ground state & 9.87 & 20.19 & 27.42 & 33.21 \\
$\mu_1$ & 0 & 1 & 3/2 & 2 \\
$\nu_{\text{ground}}$ & 0 & 1 & 3/2 & 2 \\
\hline
\end{tabular}
\end{table}

%#############################################################################
\section{Application to the Hydrogen Atom}
\label{sec:hydrogen}
%#############################################################################
For the hydrogen atom with Coulomb potential $V(r) = -e^2/(4\pi\epsilon_0 r)$, the radial Schr\"odinger equation is
\begin{equation}
\left[-\frac{\hbar^2}{2m_e}\frac{1}{r^2}\frac{d}{dr}\left(r^2\frac{d}{dr}\right) + \frac{\ell(\ell+1)\hbar^2}{2m_e r^2} - \frac{e^2}{4\pi\epsilon_0 r}\right]R = ER.
\label{eq:hydrogen_radial}
\end{equation}
With boundary conditions $R(0)$ finite and $R(\infty) \to 0$, the bound-state solutions require the principal quantum number $n = n_r + \ell + 1$ to be a positive integer, giving
\begin{equation}
E_n = -\frac{m_e e^4}{2\hbar^2(4\pi\epsilon_0)^2}\frac{1}{n^2} = -\frac{13.6\text{ eV}}{n^2}.
\label{eq:hydrogen_energy}
\end{equation}

\subsection{Origin of Integer Quantum Numbers}
The radial equation~\eqref{eq:hydrogen_radial} is mathematically well-defined for any real $\ell > -1/2$. Bound-state solutions exist for a continuum of $\ell$ values, with energy depending continuously on $\ell$. The restriction to integers comes entirely from the angular part. For the full sphere with azimuthal extent $\Phi = 2\pi$, the single-valuedness requirement $\psi(\phi + 2\pi) = \psi(\phi)$ forces $m$ to be an integer. Given integer $m$, the polar regularity condition forces $\ell$ to be an integer $\geq |m|$.

\subsection{Hydrogen Atom in a Wedge: A Conceptual Probe}
We now consider what would happen if a hydrogen atom's electron were confined to a spherical wedge with azimuthal extent $\Phi < 2\pi$, with infinite potential barriers at $\phi = 0$ and $\phi = \Phi$ but no radial confinement.\\
\textbf{A note on physical realisability.} A literal ``hydrogen atom in a wedge'' is not directly realisable: one would need infinite potential walls that affect the electron but not the proton, extending from $r = 0$ to $r = \infty$. No known mechanism creates such selective azimuthal confinement. The value of this construction is therefore \emph{conceptual}: it demonstrates that the integer angular momentum spectrum arises from boundary conditions, not from the Coulomb potential itself. The Coulomb radial quantisation mechanism works equally well with non-integer effective angular momentum; it is the angular boundary conditions that select integers in the standard case.\\
\textbf{Mathematical treatment.} With this understanding, we proceed with the mathematical analysis. The Schr\"odinger equation still separates in spherical coordinates. The radial equation remains~\eqref{eq:hydrogen_radial}, but the angular part is solved with wedge boundary conditions. The effective azimuthal quantum number is $\mu = n_\phi\pi/\Phi$, and the polar quantum number is $\nu = \mu + k$ for $k = 0, 1, 2, \ldots$.

The radial quantisation proceeds as in standard hydrogen. The bound-state condition (regularity at $r = 0$, normalisability as $r \to \infty$) gives the principal quantum number $n = n_r + \nu + 1$, where $n_r = 0, 1, 2, \ldots$ is the number of radial nodes. The energy levels become
\begin{equation}
E = -\frac{13.6\text{ eV}}{(n_r + \nu + 1)^2} = -\frac{13.6\text{ eV}}{(n_r + n_\phi\pi/\Phi + k + 1)^2}.
\label{eq:hydrogen_wedge}
\end{equation}

\textbf{Symmetry breaking.} The wedge boundary conditions break the SO(3) rotational symmetry, so $\hat{L}_z$ no longer commutes with the full Hamiltonian-plus-boundary system and $m$ ceases to be a good quantum number. More significantly, the SO(4) dynamical symmetry discovered by Pauli~\cite{Pauli1926} and Fock~\cite{Fock1935}---which explains the ``accidental'' $n^2$-fold degeneracy of hydrogen---is also broken. In standard hydrogen, states with the same principal quantum number $n$ but different $\ell$ are degenerate because they are related by the Laplace-Runge-Lenz vector, which commutes with the Hamiltonian. The wedge boundary conditions destroy this symmetry, and states with the same effective $n$ but different $\nu$ are no longer degenerate.

For a third-sphere wedge ($\Phi = 2\pi/3$), the ground state has $n_\phi = 1$, $k = 0$, $n_r = 0$, giving $\nu = 3/2$ and
\begin{equation}
E_{\text{ground}} = -\frac{13.6\text{ eV}}{(0 + 3/2 + 1)^2} = -\frac{13.6\text{ eV}}{(5/2)^2} = -2.18\text{ eV}.
\label{eq:hydrogen_wedge_ground}
\end{equation}
This energy appears nowhere in the standard hydrogen spectrum.

\subsection{Physical Analogues}
While literal ``hydrogen in a wedge'' is not realisable, the underlying physics---modified angular momentum quantisation due to geometric constraints---does appear in experimentally accessible systems: \\
\textbf{Quantum dots with shaped confinement.} Semiconductor quantum dots can be fabricated with non-circular geometries~\cite{ReimannManninen}. Shell structure and angular momentum quantisation in circular quantum dots have been observed experimentally~\cite{Tarucha1996}, and deviations from circular symmetry are known to modify the angular momentum spectrum. A wedge-shaped quantum dot would exhibit the modified spectrum derived here (without the Coulomb singularity at $r = 0$, but with similar angular structure). Such ``artificial atoms'' provide a controllable platform for studying quantum confinement effects. \\
\textbf{Cold atoms in shaped optical traps.} Ultracold atoms confined in optical dipole traps~\cite{GrimmOpticalDipole} can be placed in geometrically constrained regions. Optical lattices~\cite{BlochOpticalLattices} allow precise control of potential landscapes, and while creating a precise wedge geometry is technically challenging, approximate wedge confinement using spatial light modulators is conceivable. The absence of a Coulomb potential in such systems means they realise ``particle in a wedge'' rather than ``hydrogen in a wedge.'' \\
\textbf{Electrons on patterned surfaces.} Surface states on metals can be confined by step edges or lithographically defined boundaries. The ``quantum corral'' experiments of Crommie, Lutz, and Eigler~\cite{CrommieLutzEigler} demonstrated that electrons on Cu(111) surfaces can be confined to circular geometries and exhibit quantised standing-wave patterns. Wedge-shaped domains on surfaces would exhibit the angular momentum structure analysed here. \\
\textbf{Topological defects.} Conical geometries arising from disclinations in condensed matter systems produce effective angular deficits analogous to our wedge. Cosmic strings~\cite{Vilenkin1981, VilenkinShellard}, if they exist, create conical spacetime with a deficit angle $\delta = 8\pi G\mu$, where $\mu$ is the string's mass per unit length. Quantum fields propagating in such backgrounds experience modified angular momentum quantisation mathematically identical to our wedge model. Similar physics appears in graphene with disclination defects and in liquid crystals.

The common thread is that \emph{any} system where the effective azimuthal range differs from $2\pi$ will exhibit non-integer effective angular momentum quantum numbers. The ``hydrogen in a wedge'' calculation is a conceptual probe that reveals this general principle.

\subsection{Recovery of Integer Quantisation}
As $\Phi \to 2\pi$, the effective azimuthal quantum numbers $\mu_n = n\pi/(2\pi) = n/2$ approach half-integers for odd $n$. However, the limit $\Phi \to 2\pi$ with Dirichlet boundary conditions does not smoothly connect to the standard periodic case. This is not merely a technicality but reflects a fundamental discontinuity. The key point is that the full sphere requires a \emph{different specification of the Hilbert space domain}---periodic identification $\phi \sim \phi + 2\pi$, not Dirichlet walls. The two problems are \textbf{not connected by a continuous limit in operator topology}: the Dirichlet domain (functions vanishing at $\phi = 0$ and $\phi = \Phi$) and the periodic domain (functions satisfying $\psi(\phi + 2\pi) = \psi(\phi)$) are inequivalent subspaces of $L^2$, and consequently the spectra differ qualitatively. When $\Phi = 2\pi$ with periodic boundary conditions, the azimuthal solutions are $e^{im\phi}$ with integer $m$, not $\sin(m\phi/2)$. These are eigenfunctions of $\hat{L}_z$, and the operator $\hat{L}_z$ is self-adjoint on this domain.

In the wedge, the Hilbert space domain is changed: functions must vanish at $\phi = 0$ and $\phi = \Phi$. On this domain, the operator $\hat{L}_z = -i\hbar\partial/\partial\phi$ is not self-adjoint (it maps functions satisfying the boundary conditions to functions violating them). The symmetry generated by $\hat{L}_z$ is broken, and $m$ is no longer a good quantum number. The standard integer spectrum is recovered exactly when the correct domain (full $2\pi$ with periodic identification) is imposed. The wedge model shows that this integer spectrum follows from the specification of the Hilbert space, not from any intrinsic property of the Schr\"odinger operator or the Coulomb potential.

%#############################################################################
\section{Conclusion}
\label{sec:conclusion}
%#############################################################################
We have presented the complete solution of the Schr\"odinger equation for a particle confined to a spherical wedge, an exactly solvable constrained-domain model that exhibits spectral reorganisation under symmetry-breaking boundary conditions. The azimuthal Dirichlet conditions produce standing-wave solutions with effective quantum number $\mu = n_\phi\pi/\Phi$, which is generically non-integer. The standing-wave structure has an important consequence: the stationary states are not eigenstates of $\hat{L}_z$, and we proved explicitly that $\langle L_z \rangle = 0$ while $\Delta L_z = \hbar\mu \neq 0$. This reflects the fact that $\hat{L}_z$ is not self-adjoint on the wedge Hilbert space---it maps functions satisfying the boundary conditions to functions that violate them. The symmetry generated by $\hat{L}_z$ is broken, and the $z$-component of angular momentum becomes an observable with genuine quantum uncertainty rather than a good quantum number. The polar equation yields associated Legendre functions of non-integer order, with the Ferrers function of the second kind $Q_\nu^\mu$ excluded on integrability grounds. Square-integrability at both poles requires $\nu - \mu \in \mathbb{Z}_{\geq 0}$, establishing a hierarchy among angular solutions: sectoral states ($\nu = \mu$) satisfy the first-order highest-weight condition and exist for any $\mu > 0$, while tesseral states require integer steps $\nu = \mu + k$.

The ``hydrogen in a wedge'' construction, while not directly realisable experimentally, serves as a conceptual probe demonstrating that the integer angular momentum spectrum arises from boundary conditions, not the Coulomb potential. The Coulomb radial quantisation works equally well with non-integer $\nu$; it is the angular domain specification that selects integers. The wedge boundaries also break the SO(4) dynamical symmetry (generated by the Laplace-Runge-Lenz vector) that underlies the $n^2$ degeneracy of standard hydrogen. Physical systems exhibiting related physics include quantum dots with non-circular confinement, cold atoms in shaped optical traps, and electrons near topological defects. The model clarifies the distinct roles of two constraints in the standard quantisation of orbital angular momentum. Single-valuedness under $2\pi$ rotation (arising from the periodic identification $\phi \sim \phi + 2\pi$) selects integer $m$. Regularity at both poles (arising from the structure of the associated Legendre equation) then selects integer $\ell \geq |m|$. Crucially, the Dirichlet and periodic problems are not connected by a continuous limit in operator topology---they define inequivalent Hilbert space domains with qualitatively different spectra.

We emphasise that this paper presents a pedagogically useful exactly solvable model, not a claim of new physics. The mathematical tools---separation of variables, associated Legendre functions, spherical Bessel functions---are entirely standard. The value of the model lies in the explicit, algebraically complete demonstration of how the specification of the operator domain affects angular momentum quantisation, providing concrete illustration of principles that can otherwise remain abstract.

\end{document}